\documentclass[10pt]{article}

\usepackage[margin=0.82in]{geometry}
\usepackage{graphicx}
\usepackage{amsmath,amssymb}
\usepackage{authblk}
\usepackage[backend=biber,style=numeric-comp,sorting=none,giveninits=true,maxbibnames=3,minbibnames=3,doi=false,url=false,eprint=false]{biblatex}
\DeclareSourcemap{\maps[datatype=bibtex]{\map{\step[fieldset=abstract,null]}}}
\usepackage{jabbrv}
\usepackage{xcolor}
\usepackage{microtype}
\usepackage{hyperref}
\usepackage{caption}
\usepackage{subcaption}
\usepackage{booktabs}
\usepackage{bm}
\usepackage{titlesec}
\titleformat{\section}{\normalfont\large\bfseries}{\thesection}{0.6em}{}

\definecolor{deepgreen}{RGB}{0, 110, 50}

\hypersetup{colorlinks=true,citecolor=blue,linkcolor=blue,urlcolor=blue}

\title{Giant thermal transduction via feedback-phase engineering in a thin-film lithium-niobate opto-electro-mechanical oscillator}

\author[1,2,$\dagger$]{Ran Yin}
\author[1,2,3,$\dagger$]{Yue Yu}
\author[1,2]{Ian Christen}
\author[1,2,3,*]{Mengjie Yu}

\affil[1]{Department of Electrical Engineering and Computer Sciences, University of California, Berkeley, Berkeley, CA 94720, USA}
\affil[2]{Ming Hsieh Department of Electrical and Computer Engineering, University of Southern California, Los Angeles, CA 90089, USA}
\affil[3]{Materials Sciences Division, Lawrence Berkeley National Laboratory, Berkeley, CA, USA}

\affil[$\dagger$]{These authors contributed equally to this work.}
\affil[*]{Corresponding author. Email: mengjie.yu@berkeley.edu}

\begin{document}
\begin{refsection} 
\twocolumn[{%
\begin{@twocolumnfalse}
\maketitle

\begin{abstract}

Conventional resonant MEMS thermometry typically relies on the intrinsic temperature coefficient of frequency (TCF) of mechanical resonances. Here we show that embedding a suspended thin-film lithium-niobate mechanical resonator in an opto-electro-mechanical (OEM) feedback loop enhances the intrinsic coefficient by two orders of magnitude. By using laser--cavity detuning to control the loop's round-trip phase, and thus the Barkhausen-selected oscillation frequency, we engineer an effective TCF of $+1.24\times10^{4}$~ppm/K: a 100-fold enhancement and a sign reversal relative to its intrinsic TCF of $-124$~ppm/K. Temperature-dependent and open-loop phase measurements confirm that this giant response originates from feedback phase selection rather than from the passive resonator itself. This interpretation is further supported by the independently measured closed-loop oscillation frequencies and a detuning-dependent delay model. The same architecture also supports 2:1 subharmonic injection locking, yielding sub-6~Hz linewidths, reference-limited frequency stability, and tunable operation from $76.71$ to $76.82$~MHz. These results establish feedback-phase engineering as a general, post-fabrication route to both giant thermal transduction and low-noise yet tunable mechanical oscillation.

\end{abstract}

\vspace{0.8em}
\end{@twocolumnfalse}
}]

\section{Introduction}    
Microelectromechanical systems (MEMS), owing to their small footprint, low effective mass, and compatibility with chip-scale fabrication, provide compact and versatile platforms for sensing force, acceleration, pressure, mass, and temperature \cite{Sansa2020_mass_sens, Gavartin2012_force_sens, Kippenberg_optomechanics_cavity, Anetsberger2009_string_displacement_sens, Kim2017_torch_sens, Zhang17_femtogram_water_mass_sens_future, Basir19_ultrasound}. Among these applications, MEMS thermometry is used for localized temperature monitoring and thermal management in integrated electronic and photonic systems, and for compensating temperature-induced frequency drift in resonant sensors, oscillators, and timing references \cite{MEMs_tempt_sens_calibration, sui2025review, Kaya_dual_res_tem_sens, tu2022coupled_out_of_plane_vib_1000ppmK, Mussi_MEMs_Tem_cali}. In resonant MEMS thermometers, temperature-dependent changes in the effective stiffness and geometry of the device shift its mechanical resonance frequency, which can be tracked with high precision when the resonance quality factor ($Q$) is large. These thermometers offer scalability to large-scale sensor arrays \cite{Zhang2013_torsional_freqshift_IR}. The temperature responsivity is characterized by the temperature coefficient of frequency (TCF), defined as the fractional resonance-frequency shift per unit temperature change. Conventional MEMS resonators typically exhibit TCF magnitudes ranging from tens to hundreds of ppm/K \cite{Qian2016_MEMs_GaN_IRsensor, sui2025review, Blaikie2019_graphene_bolometer, Hui16plasma_pe_IRsens, Markus_electromechanical_IRsens, 12um_IR_sens_electromechanical, ZhenyunQian_AlN_IRsens}, and various approaches---thermally induced stress engineering \cite{clamp_4537ppmK, clamp_-3550ppmK, zou2021compact_clamp, graphene_-3180ppmK}, temperature-responsive materials \cite{polymer_8e4ppmK}, and alternative frequency-readout schemes \cite{BEATING_SIGNAL_9929ppmK, comb_space_-11481ppmK, tu2022coupled_out_of_plane_vib_1000ppmK, dual_mode_1480ppmK}---have pushed TCF magnitudes up to several thousand ppm/K. The current record is held by shape-memory polymer resonators, which reach TCF magnitudes of approximately $8\times10^4$~ppm/K near their glass-transition temperature, albeit only over a narrow temperature window and with strong mechanical damping ($Q_m\approx2$) \cite{polymer_8e4ppmK}.

Precision sensing based on this frequency-shift modality further benefits from self-sustained mechanical oscillation with a narrow spectral linewidth and high signal-to-noise ratio (SNR), which can be achieved through regenerative feedback that compensates for intrinsic mechanical dissipation via either electromechanical \cite{feng2008nems} or optomechanical transduction \cite{Kippenberg_optomechanics_cavity}. In optomechanical systems based on piezoelectric materials, the optically transduced mechanical signal can be photodetected, electrically amplified, and fed back to the resonator through piezoelectric actuation, with all-optical frequency-shift readout avoiding electronic and readout-circuit noise. This opto-electro-mechanical (OEM) feedback architecture provides regenerative gain to sustain mechanical oscillation and has been explored to reduce phase noise relative to the passive mechanical resonance \cite{horvath2023closedloop}. In addition, optoelectronic oscillators (OEOs), which do not incorporate a mechanical resonator, have demonstrated excellent temperature-sensing performance \cite{OEO_sens_generally, OEO_tem_sens_1.26GHzK, OEO_tem_sens_10GHzK, OEO_tem_sens_1MHzK, OEO_tem_sens_2.7MHzK, OEO_tem_sens_3.7MHzK, OEO_tem_sens_44kHzK, OEO_tem_sens_7.7GHzK, OEO_tem_sens_997kHzK, OEO_LN_intensity, fiber_delay_temperature_sense_OEO_venier, delay_OEO_fiber_strain}. Although OEO and OEM systems both employ regenerative feedback, their transduction pathways are distinct: in conventional OEO architectures, an electro-optic (EO) modulator transfers the RF feedback signal onto an optical carrier, which is subsequently converted back into an electrical signal through photodetection \cite{OEO_tem_sens_7.7GHzK, OEO_tem_sens_10GHzK}.

In contrast, the OEM oscillator investigated here relies on the mechanical motion of a suspended, $40$~$\mu$m-diameter thin-film lithium-niobate (TFLN) resonator that co-integrates the optical and mechanical modes, naturally providing a narrow filter response set by the mechanical $Q$ to generate the optical modulation, while the same resonator also provides piezoelectric actuation in response to the electrical feedback signal. This mechanically mediated feedback eliminates the need for a separate EO modulator to sustain free-running oscillation and introduces a mechanical degree of freedom into the frequency-selection process. TFLN is particularly well suited to this architecture because it combines strong piezoelectric coupling with low-loss optical confinement \cite{wang2024lnpiezo, zhang2018ultrahighq}. 

Furthermore, we investigate how the resulting regenerative feedback modifies the temperature dependence of the oscillation frequency. Closing the OEM feedback loop increases the measured mechanical SNR from approximately $10$ to $58.5$~dB and yields an oscillation linewidth of $4.5$~kHz. 
The oscillation frequency exhibits an effective TCF of $+1.24\times10^{4}$~ppm/K---two orders of magnitude larger than, and opposite in sign to, the passive-mode TCF of $-124$~ppm/K for this resonator platform \cite{uncooledlownoisethinfilmoptomechanical}. We discover that the high intrinsic optical $Q$, on the order of $10^{6}$, increases the temperature-transduction sensitivity by modifying the Barkhausen phase condition satisfied by the mechanical oscillation via the laser--cavity detuning. This mechanism is validated by both open-loop and closed-loop measurements and agrees well with a numerical model. Finally, we demonstrate injection locking of the OEM oscillator to an external RF reference, yielding a resolution-limited linewidth below $6$~Hz over a locking range from $76.71$ to $76.82$~MHz, establishing a complementary route to frequency-stabilized, externally referenced mechanical oscillation.

\begin{figure*}[!t]
    \centering
    \includegraphics[width=0.97\textwidth]{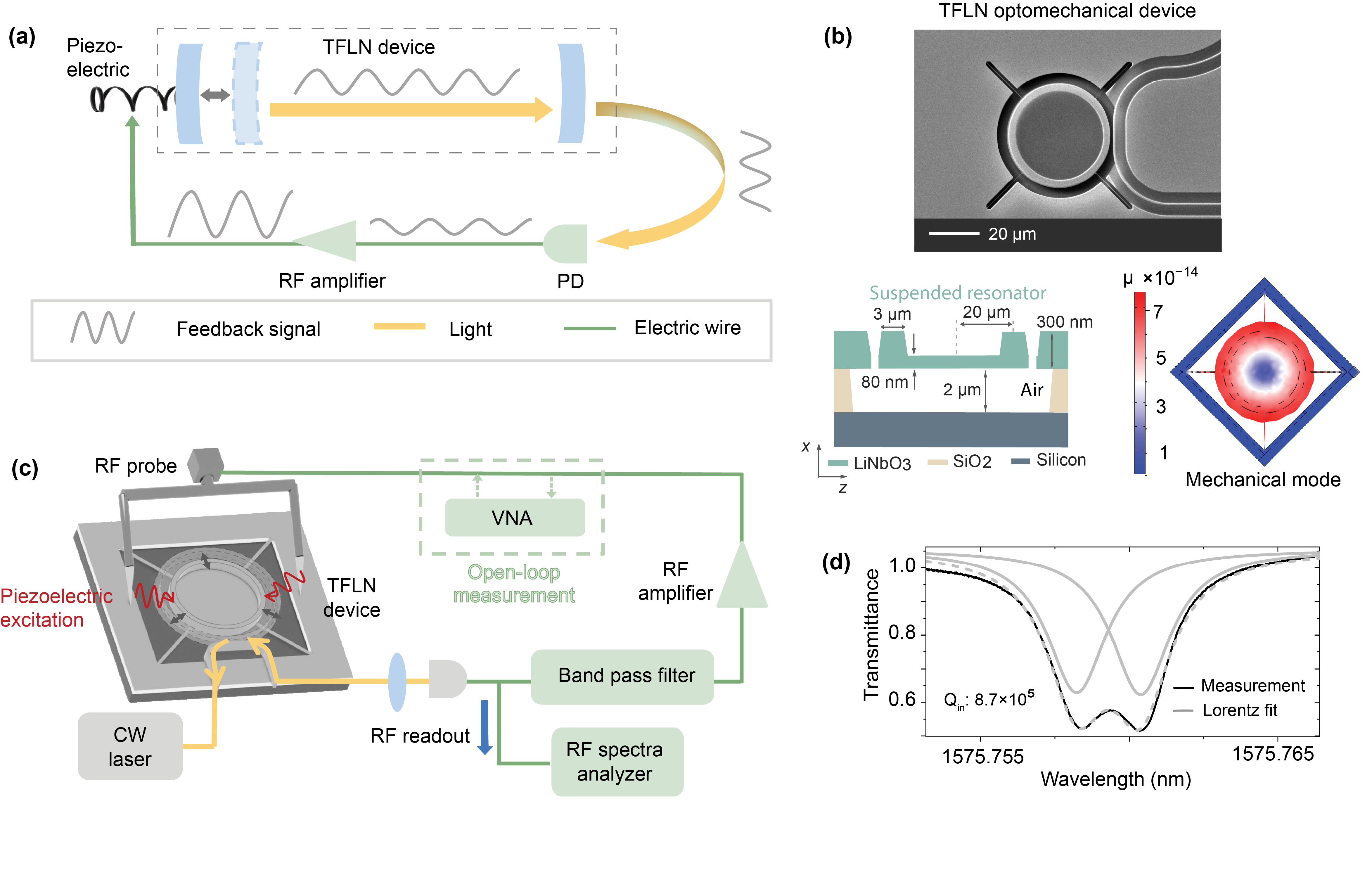}
    \caption{\textbf{Opto-electro-mechanical feedback oscillator based on a suspended TFLN device.}
    (a) Conceptual OEM feedback loop, illustrating the conversion between mechanical motion, optically transduced signals, and electrically amplified piezoelectric feedback.
    (b) Suspended TFLN resonator: scanning electron micrograph (top), cross-sectional structure showing the LiNbO$_3$, SiO$_2$, silicon, and air regions (bottom left), and simulated fundamental breathing-mode displacement profile (bottom right).
    (c) Experimental configuration. The photodetected mechanically modulated signal is filtered, amplified, and applied back to the device through an RF probe. A vector network analyzer (VNA) can be optionally inserted for open-loop $S_{21}$ spectra measurement; an RF spectrum analyzer records the closed-loop response. The band-pass filter has a specified range of $67$--$88$~MHz, and the RF amplifier has a bandwidth extending to $2$~GHz, with flat gain spectra across the mechanical mode range.
    (d) Optical transmission spectrum of the suspended resonator (black) and Lorentzian-fitted resonance doublet (gray). The resonance splitting arises from backscattering-induced coupling between clockwise and counterclockwise optical modes, with the supporting anchors along the microring perimeter contributing to the enhanced backscattering.}
    \label{fig:setup}
\end{figure*}

\section{Opto-electro-mechanical feedback oscillator}

 As illustrated conceptually in Figure~\ref{fig:setup}(a), mechanical displacement shifts the optical cavity resonance and thereby modulates the transmitted optical field. A photodetector converts this optical modulation into an RF signal, which is amplified and applied back onto the device. Through the piezoelectric response of TFLN, the returned electrical signal drives the mechanical motion again, completing a feedback cycle between the mechanical, optical, and electrical domains. When the returned mechanical drive reinforces the existing motion with sufficient gain, the feedback can compensate the mechanical dissipation and lead to oscillation.

The suspended TFLN optomechanical resonator, a disk mechanical resonator with an optical whispering gallery mode confined in the ring around the edge, is shown in Figure~\ref{fig:setup}(b). The cross-sectional schematic illustrates the LiNbO$_3$ layer suspended above the silicon substrate while the SiO$_2$ buffer is locally removed beneath the resonator to mechanically isolate the system. The simulated fundamental breathing mode exhibits radial deformation of the microring and thus modulation of the optical cavity resonance through optomechanical coupling. Meanwhile, an RF probe contacting the LN surface adjacent to the suspended region applies an oscillating electric field across the resonator to drive mechanical motion through the piezoelectric response intrinsic to LiNbO$_3$ (Figure~\ref{fig:setup}(c)). By integrating optical readout and piezoelectric actuation within a single suspended resonator, this device provides the essential mechanical interface for the regenerative OEM feedback loop. To realize the complete feedback loop experimentally, a continuous-wave laser is coupled to the optical resonance of the suspended device with intrinsic $Q$ of $0.87\times10^{6}$ (Figure~\ref{fig:setup}(d)). The mechanically modulated transmitted light is detected, filtered, electrically amplified, and then fed back to the resonator through the RF probe. The band-pass filter with a passband of approximately 67--88~MHz restricts the feedback to the frequency range containing the fundamental breathing-mode response.

Self-sustained oscillation in a regenerative feedback loop is governed by the Barkhausen condition, which requires the signal to reproduce itself in amplitude and phase after one complete round trip. We express the mechanical signal after one round trip as
\begin{equation}
    E_{\rm mech}^{(n+1)}
    =G_{\rm rt}e^{i\Phi_{\rm rt}}E_{\rm mech}^{(n)},
    \label{eq:roundtrip}
\end{equation}
where $G_{\rm rt}$ and $\Phi_{\rm rt}$ denote the round-trip magnitude gain and phase, respectively. For steady-state free-running oscillation, the Barkhausen condition requires unity round-trip gain ($G_{\rm rt}=1$) and an accumulated phase equal to an integer multiple of $2\pi$ ($\Phi_{\rm rt}=2\pi m$), where $m$ is an integer~\cite{feng2008nems}.

\begin{figure}[tbp]
    \centering
    \includegraphics[width=0.98\columnwidth]{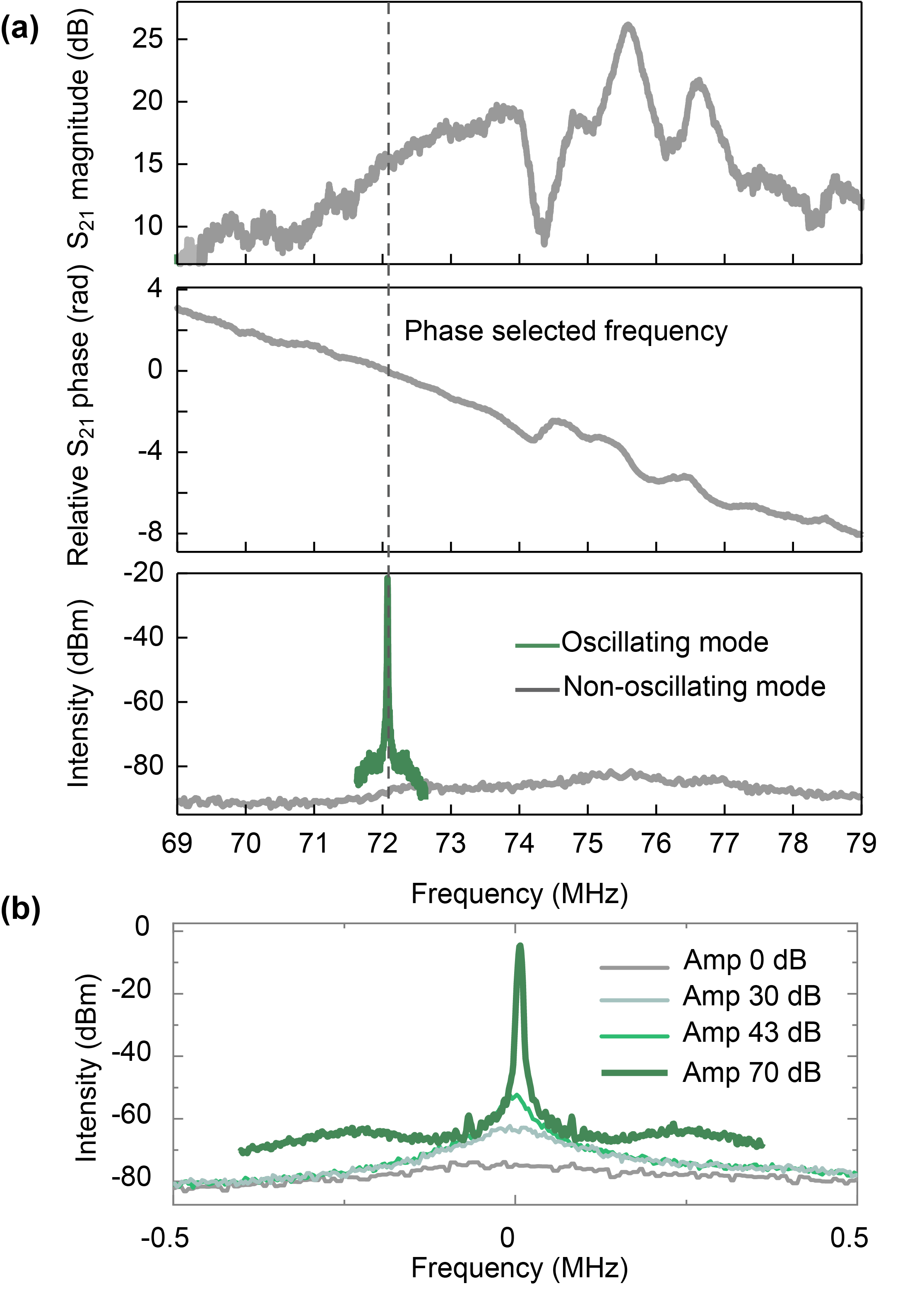}
    \caption{\textbf{Open-loop characterization and onset of mechanical self-oscillation.}
    (a) Measured open-loop $S_{21}$ magnitude (top) and relative phase (middle), together with the passive and closed-loop mechanical RF spectra (bottom). The vertical dashed line denotes the measured oscillation frequency under closed loop operation, which is used to establish the experimental phase-selection reference. The passive mechanical response has an SNR of approximately $10$~dB and a linewidth of $1.42$~MHz; feedback oscillation yields an SNR of $58.5$~dB and a linewidth of $4.5$~kHz.
    (b) RF spectra of the mechanical mode measured with amplifier gains of $0$, $30$, $43$, and $70$~dB, showing the increased SNR and narrowed linewidth of the optomechanical signal and the stable self-sustained oscillation at the applied gain of $70$~dB.}
    \label{fig:openloop}
\end{figure}

To characterize these conditions experimentally, we insert a VNA into the feedback path to measure the open-loop $S_{21}$ magnitude and phase responses. The magnitude spectrum, shown in the upper trace of Figure~\ref{fig:openloop}(a), exhibits several peaks corresponding to the mechanical resonances of the suspended microresonator, where the mechanically mediated feedback response is enhanced. A comparison with the passive mechanical spectrum is provided in Supplementary Information, Section~1. The measured open-loop gain exceeds unity over a broad frequency range encompassing these resonances, indicating that sufficient gain is available to sustain oscillation at multiple frequencies. Consequently, the accumulated loop phase plays a critical role in selecting the final oscillation frequency.

The corresponding open-loop phase response is shown in the middle trace of Figure~\ref{fig:openloop}(a). We establish the experimental phase-selection reference to be the phase measured at the independently observed closed-loop oscillation frequency, indicated by the vertical dashed line (Figure~\ref{fig:openloop}(a) lower panel). The phase spectrum is then expressed relative to this reference, allowing the phase-selected frequency to be identified from the open-loop measurements. As shown in the power density spectra in Figure~\ref{fig:openloop}(a), the passive mechanical response exhibits multiple resonances, with an SNR of approximately $10$~dB and a linewidth of $1.42$~MHz. Closing the feedback loop produces self-sustained oscillation near $72.1$~MHz, increasing the SNR to $58.5$~dB and narrowing the linewidth to $4.5$~kHz, thereby providing a well-defined spectral feature for frequency-domain readout. Figure~\ref{fig:openloop}(b) further shows the measured optomechanical signal at applied RF gains ranging from $0$ to $70$~dB: the spectral linewidth narrows as the increasing gain reduces the effective closed-loop loss, finally leading to a stable self-sustained oscillation at a gain of $70$~dB.

\section{Two orders of magnitude higher TCF and the mechanism}
\begin{figure}[tbp]
    \centering
    \includegraphics[width=0.98\columnwidth]{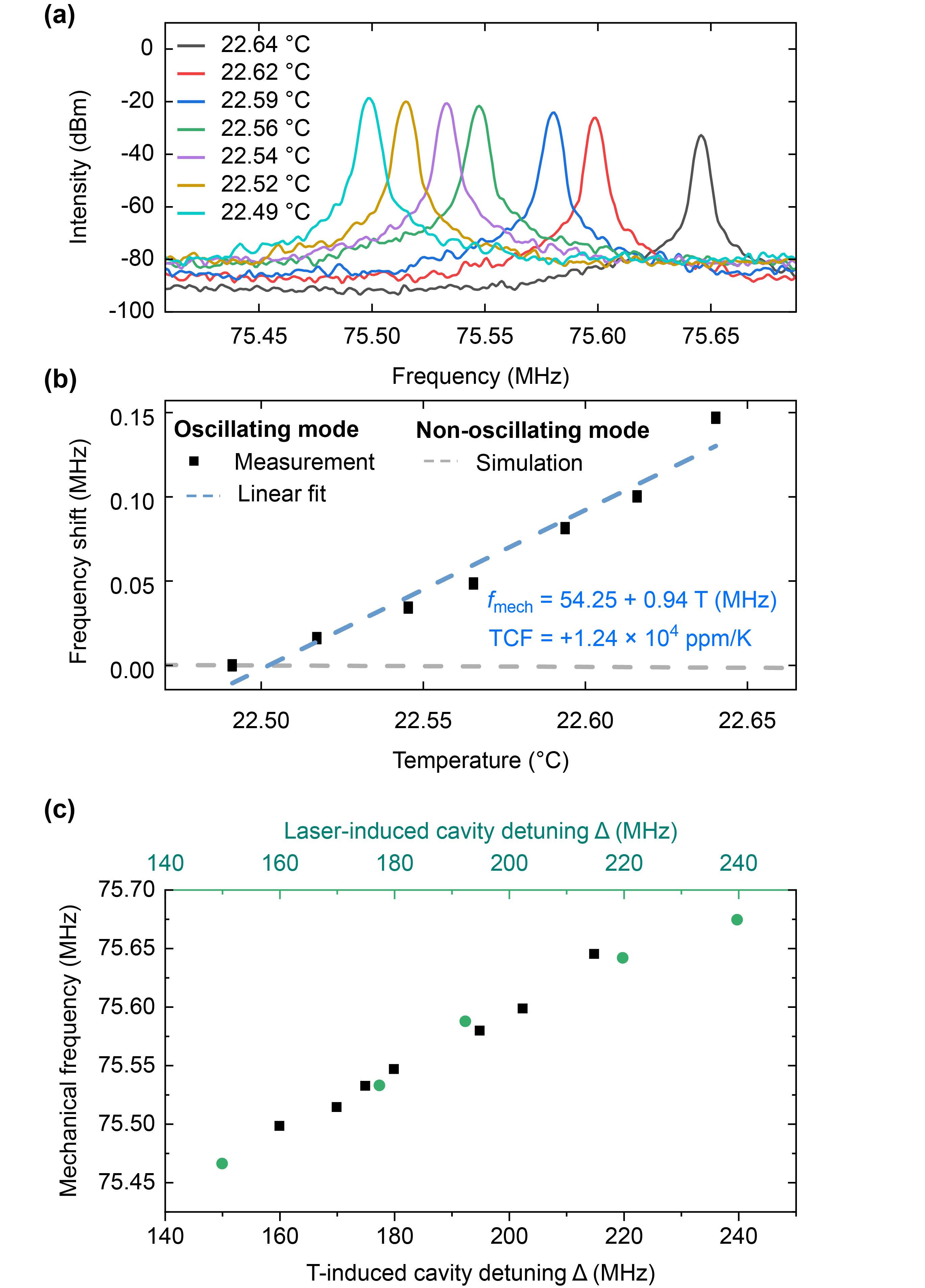}
    \caption{\textbf{Giant effective temperature coefficient and its dependence on laser--cavity detuning.}
    (a) Closed-loop RF spectra measured at different device temperatures.
    (b) Oscillation-frequency shift versus temperature. The linear fit gives $df_{\rm osc}/dT\approx0.94$~MHz/$^{\circ}$C and an effective TCF of $+1.24\times10^{4}$~ppm/K. The gray dashed line represents the simulated passive-mode TCF of $-124$~ppm/K, consistent with the experimentally measured value for a suspended device of similar geometry reported in Ref.~\cite{uncooledlownoisethinfilmoptomechanical}. The temperature-dependent behavior beyond this high-TCF operating range is discussed in Supplementary Information, Section~2.
    (c) Oscillation frequency as a function of laser--cavity detuning produced by temperature tuning at fixed laser frequency (black squares, lower detuning axis) and laser-frequency tuning at fixed temperature (green circles, upper detuning axis). Both axes represent laser--cavity detuning and share the same tick positions, distinguishing how the detuning was varied.}
    \label{fig:tcf}
\end{figure}

The narrow linewidth and high SNR of the OEM oscillation signal make it well suited for sensing applications. We therefore investigate its potential for temperature sensing. As the device temperature increases, the oscillation frequency exhibits a pronounced shift (Figure~\ref{fig:tcf}(a)). Over the measured temperature range of approximately $22.49$--$22.64~^{\circ}\mathrm{C}$, the oscillation frequency changes approximately linearly with temperature, yielding a slope of $0.94$~MHz/$^{\circ}$C with a center frequency of about $75.5$~MHz (Figure~\ref{fig:tcf}(b)). This corresponds to a TCF of approximately $+1.24\times10^{4}$~ppm/K. For comparison, the gray dashed line in Figure~\ref{fig:tcf}(b) represents the temperature dependence of the passive mechanical resonance, reproduced from the Supplementary Information of Ref.~\cite{uncooledlownoisethinfilmoptomechanical}, which exhibits a TCF of $-124$~ppm/K. The closed-loop OEM oscillator thus exhibits a 100-fold enhancement in TCF magnitude, surprisingly with a reversal in sign relative to the passive mechanical mode, suggesting a mechanism distinct from the temperature-dependent stiffness and geometry effects that govern the passive TCF. To investigate this mechanism, we consider thermal effects beyond these elastic and geometric contributions. Specifically, the thermo-optic effect shifts the resonance frequency of the optical resonator and thus changes the laser--cavity detuning $\Delta$. To determine its contribution, we compare the oscillation-frequency dependence on $\Delta$ obtained through independent temperature tuning and laser-frequency tuning. Figure~\ref{fig:tcf}(c) shows the oscillation frequency as a function of $\Delta$ when the device temperature is varied at fixed laser frequency (black squares), as well as when the laser frequency is varied while holding the device temperature fixed (green circles). The two measurements exhibit nearly the same frequency--detuning trend, identifying the laser--cavity detuning as the dominant intermediate variable governing the giant thermal transduction of the mechanical oscillation frequency.

\begin{figure*}[!t]
    \centering
    \includegraphics[width=0.97\textwidth]{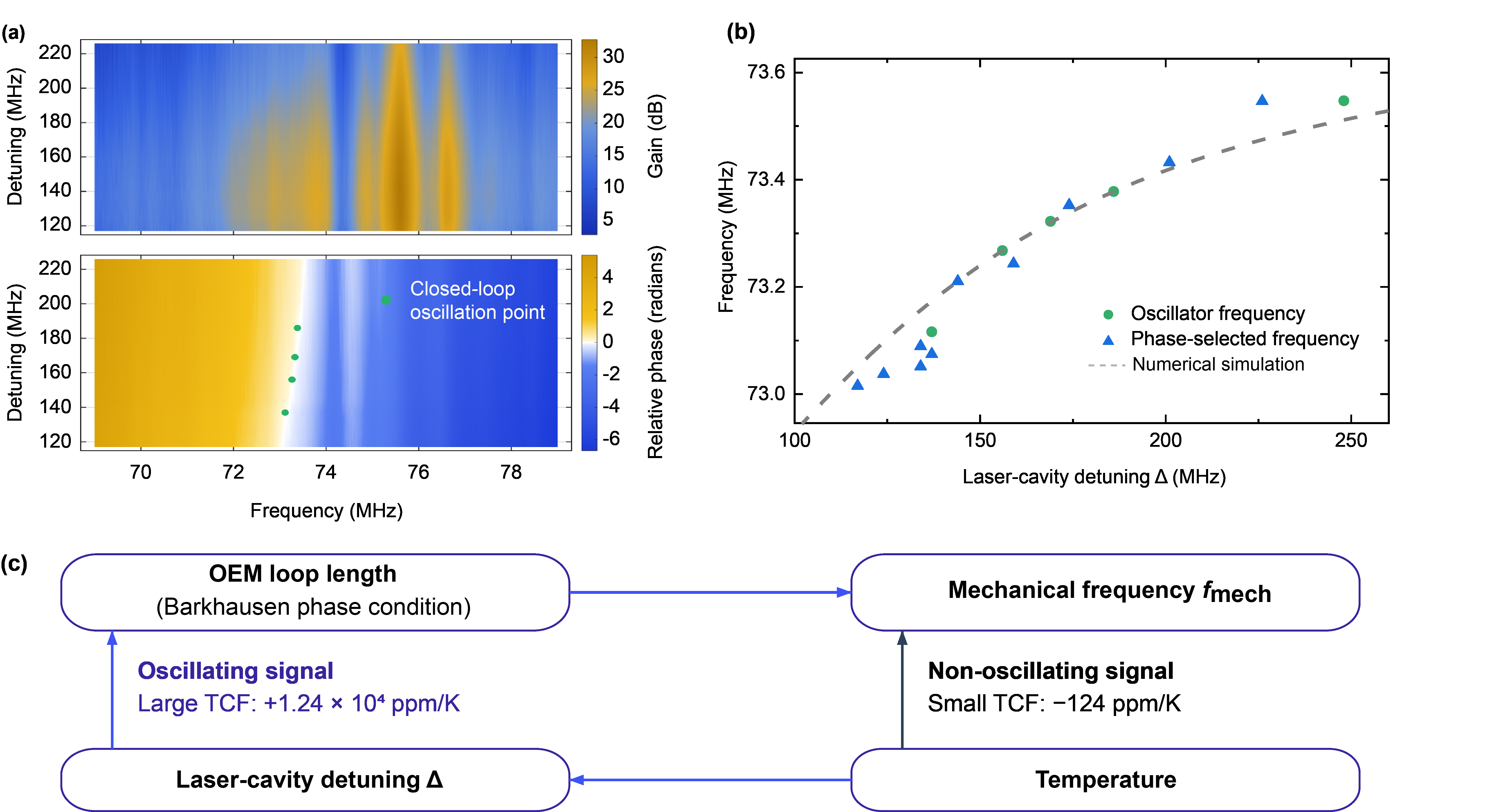}
    \caption{\textbf{Detuning-dependent phase selection and effective cavity-induced delay underlying the enhanced TCF.}
    (a) Measured open-loop $S_{21}$ magnitude (top) and relative phase (bottom) spectra under different laser--cavity detunings. The dominant magnitude features exhibit negligible frequency shift with detuning, whereas the relative phase-selection contour shifts noticeably and tracks the closed-loop oscillation frequency, identifying the detuning-dependent loop phase as the principal frequency-selection mechanism, given the broadband, positive open-loop gain.
    (b) Independently measured closed-loop oscillation frequencies (green circles), phase-selected frequencies extracted from open-loop VNA measurements (blue triangles), and the numerical simulation of the effective cavity-induced delay (gray dashed curve), showing agreement in their detuning-dependent frequency trends.
    (c) Temperature-to-frequency transduction pathways: a small direct passive mechanical TCF ($-124$~ppm/K) and a dominant detuning-mediated shift of the OEM phase-selection condition yielding a giant positive effective TCF ($+1.24\times10^{4}$~ppm/K).}
    \label{fig:phase}
\end{figure*}

We next investigate the underlying cause by examining the Barkhausen condition via an open-loop measurement using a VNA, as indicated in Figure~\ref{fig:setup}(c). Figure~\ref{fig:phase}(a) plots the open-loop $S_{21}$ magnitude and phase spectra as functions of laser--cavity detuning $\Delta$. While the prominent features of the magnitude response exhibit little frequency shift with detuning, the frequency at which the phase-selection condition is satisfied shifts systematically with detuning, as indicated by the inclined white constant-phase contour in the relative $S_{21}$ phase map. The independently measured closed-loop oscillation frequency closely follows this contour across different detunings, demonstrating that the observed oscillation-frequency tuning is governed predominantly by the detuning-dependent phase-selection condition.

To quantitatively describe this detuning-dependent phase response, we consider the intracavity energy modulation induced by mechanical motion. Mechanical displacement modulates the optical resonance frequency and, consequently, the laser--cavity detuning. Eq.~\eqref{eq:GU} shows the complex transfer function $H_U\equiv\delta U/\delta\Delta$, relating the mechanically induced detuning modulation $\delta\Delta$ to the resulting intracavity energy modulation $\delta U$:
\begin{equation}
\begin{aligned}
H_U(\Delta,f_{\rm osc})
&=\frac{\delta U}{\delta\Delta}\\
&=\frac{-2\Delta U_0(\Delta)}
{(\kappa/2-i2\pi f_{\rm osc})^2+\Delta^2},
\end{aligned}
\label{eq:GU}
\end{equation}
where $U_0(\Delta)$ is the steady-state intracavity optical energy, 
$\Delta=\omega_L-\omega_c$ denotes the laser--cavity detuning, $f_{\rm osc}$ is the mechanical oscillation frequency, and $\kappa=2\pi\Gamma$ is the optical energy-decay rate, with $\Gamma$ denoting the optical linewidth (FWHM). Equation~\eqref{eq:GU} describes the dependence of both the amplitude and phase of the intracavity energy modulation on detuning and oscillation frequency. The detailed derivation is provided in Supplementary Information, Section~3.

To quantify the cavity-induced phase variation, we select an experimentally measured VNA phase-selected frequency point, $(\Delta=\Delta_{\rm ref},f_{\rm osc}=f_{\rm ref})$, as the calibration reference. Taking its phase as the relative reference, we define the cavity-induced phase variation as

\begin{equation}
\phi_{\rm res}(\Delta,f_{\rm osc})
=\arg\left[
\frac{H_U(\Delta,f_{\rm osc})}
{H_U(\Delta_{\rm ref},f_{\rm ref})}
\right],
\label{eq:cavityphase}
\end{equation}

such that $\phi_{\rm res}(\Delta_{\rm ref},f_{\rm ref})=0$. The corresponding relative phase-equivalent time delay introduced by the cavity transduction response is

\begin{equation}
\tau_{\rm res}(\Delta,f_{\rm osc})
=\frac{\phi_{\rm res}(\Delta,f_{\rm osc})}
{2\pi f_{\rm osc}}.
\label{eq:cavitydelay}
\end{equation}


We express the total effective loop delay as the sum of the cavity-induced contribution $\tau_{\rm res}$ and a detuning-independent effective background delay, $\tau_{\rm ext}$. From the measured open-loop $S_{21}$ phase spectrum, we obtain a loop free spectral range (FSR) of $6.4$~MHz. Together with the oscillation frequency at the calibration point, this gives an effective feedback-loop phase-cycle index of $m=11$ and offset frequency of $f_{\rm off}=2.8$~MHz under the frequency-offset convention detailed in Supplementary Information, Section~4. Then, one can numerically solve for the oscillation frequency at each detuning using

\begin{equation}
\begin{aligned}
f_{\rm osc}(\Delta)
&=m\,\mathrm{FSR}_{\rm loop}+f_{\rm off}\\
&=\frac{m}{
\tau_{\rm ext}+
\tau_{\rm res}(\Delta,f_{\rm osc})
}+f_{\rm off}.
\end{aligned}
\label{eq:finalmodel}
\end{equation}

The numerical results, plotted in Figure~\ref{fig:phase}(b), agree well with both the VNA phase-selected frequencies and the independently measured closed-loop oscillation frequencies, capturing their nonlinear detuning dependence and establishing the quantitative link between laser--cavity detuning and the OEM oscillation frequency.

Together, these experiments and the model support the temperature-response pathway summarized in Figure~\ref{fig:phase}(c). Temperature changes the laser--cavity detuning, which modifies the effective optical delay of the mechanically modulated optical signal and thus the phase response of the OEM loop, shifting the Barkhausen phase-selection condition and consequently the oscillation frequency. This pathway accounts for the large positive effective TCF.

\section{External injection locking and frequency stabilization}

\begin{figure*}[!t]
    \centering
    \includegraphics[width=0.97\textwidth]{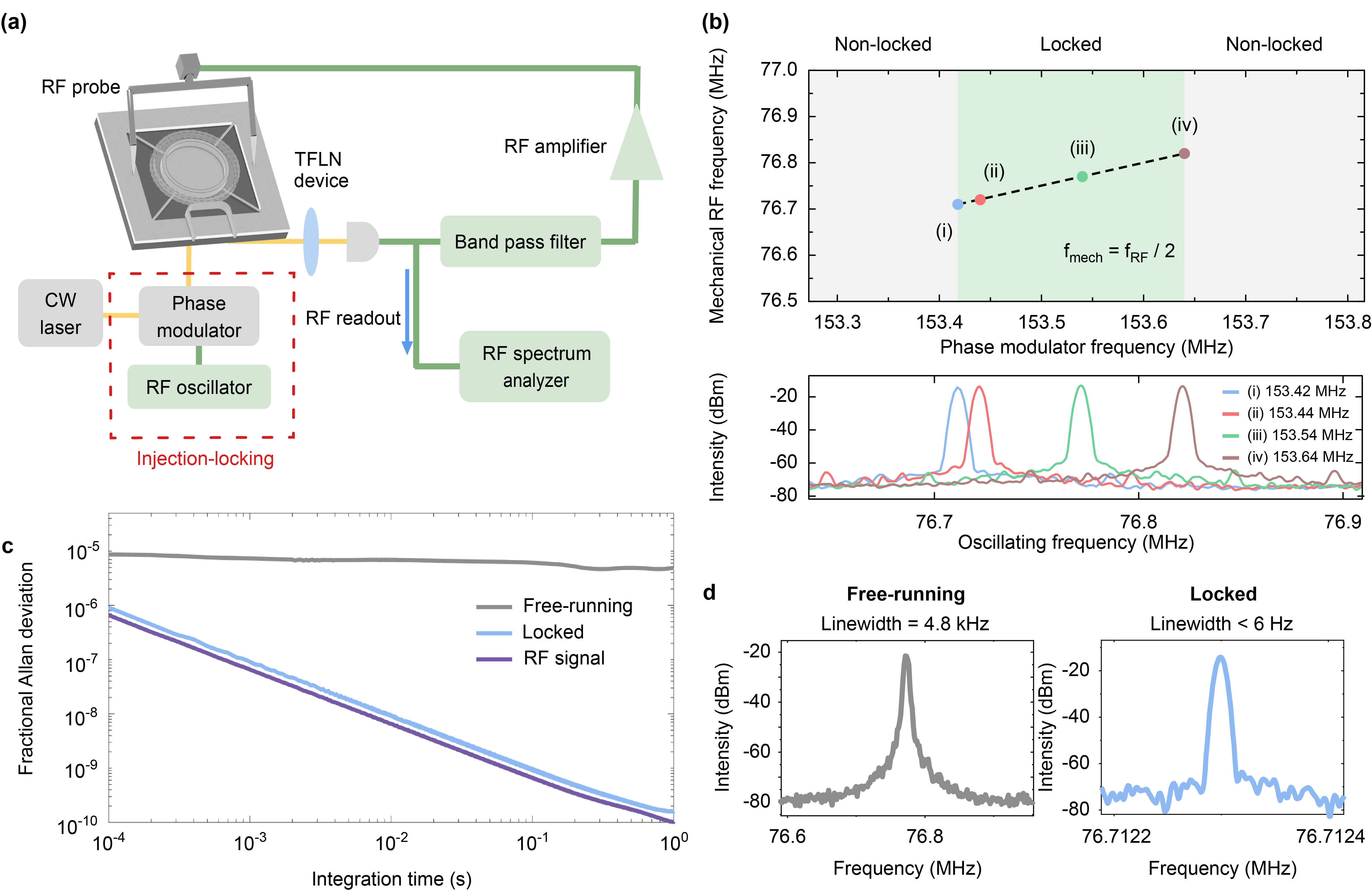}
    \caption{\textbf{Subharmonic injection locking and oscillation-frequency stabilization.}
    (a) Experimental configuration. An electro-optic phase modulator driven by an external RF source is inserted into the optical path of the OEM feedback oscillator.
    (b) RF spectra as a function of external RF frequency, showing locked and non-locked regimes. The dashed black line indicates the subharmonic relation $f_{\rm osc}=f_{\rm drive}/2$; the non-locked regions contain multifrequency mechanical responses, detailed in Supplementary Information, Section~5. Within the observed locking range, the oscillation is tuned by the applied RF frequency.
    (c) Fractional Allan deviations of the free-running oscillator, injection-locked oscillator, and external RF reference. The injection-locked oscillator approaches the reference stability over the measured averaging-time range.
    (d) Free-running mechanical RF spectrum (top; linewidth approximately $4.8$~kHz) and injection-locked spectrum (bottom; resolution-limited linewidth below $6$~Hz).}
    \label{fig:locking}
\end{figure*}

While temperature sensing exploits the environmental sensitivity of the free-running OEM oscillation frequency, the same feedback loop can also provide a stable mechanical oscillation signal through injection locking, a technique previously demonstrated to suppress frequency fluctuations and narrow oscillation linewidths in optomechanical and phonon-lasing systems \cite{phonon_laser_lock_electrical_ring, optical_enhance_harmonic_mode_lock, subhertz_lock, lock_in_coherent_phonon_source}. Figure~\ref{fig:locking}(a) illustrates the experimental configuration, in which an electro-optic phase modulator (PM) is inserted into the optical path while the rest of the OEM feedback loop remains unchanged. The PM is driven at a frequency near twice the free-running oscillation frequency, $f_{\rm drive}\approx 2f_{\rm free}$, allowing the external modulation tone to be spectrally separated from the mechanical oscillation in the photodetected RF signal.

In the absence of external modulation, the free-running oscillation frequency $f_{\rm free}$ is determined by the feedback-loop phase-selection condition. When $f_{\rm drive}/2$ falls sufficiently close to $f_{\rm free}$, the external modulation entrains the oscillator, establishing $2{:}1$ subharmonic injection locking with $f_{\rm osc}=f_{\rm drive}/2$. In this regime, the oscillation frequency is injection locked to the external drive. As the applied modulation frequency is varied from $153.42$ to $153.64$~MHz, the mechanical oscillation follows half the drive frequency, spanning $76.71$ to $76.82$~MHz while maintaining a narrow spectral response (Figure~\ref{fig:locking}(b)). The black dashed line indicates the expected $2{:}1$ subharmonic relation, $f_{\rm osc}=f_{\rm drive}/2$, distinguishing the injection-locked response from the unlocked regime. When the external modulation frequency is tuned farther away from twice the free-running frequency, stable single-frequency locking is no longer observed, and multiple mechanical oscillation peaks emerge, as shown in Supplementary Information, Section~5.

To further evaluate the frequency stability achieved through injection locking, we measure the fractional Allan deviation of the mechanical oscillation, as shown in Figure~\ref{fig:locking}(c). Injection locking substantially suppresses fractional frequency fluctuations, with the Allan deviation approaching that of the external RF reference over the measured averaging-time range of $10^{-4}$ to $1$~s. This behavior indicates that the frequency stability of the locked oscillator is strongly influenced by the external reference, suggesting that further improvements may be achievable using a lower-noise RF source. The RF spectra of the oscillating signal also demonstrate significant linewidth reduction in Figure~\ref{fig:locking}(d). The free-running oscillation exhibits a linewidth of approximately $4.8$~kHz, whereas the injection-locked signal becomes resolution limited, with a measured linewidth below $6$~Hz.

\section{Discussion and conclusion}

To our knowledge, this work demonstrates the first OEM feedback oscillator on a thin-film lithium-niobate platform, establishing feedback-induced enhancement of the TCF via optical intracavity delay as a distinct mechanism for shaping the temperature response of a regenerative mechanical oscillator. This combination of giant responsivity and narrow, high-SNR linewidth yields a linewidth-normalized figure of merit $\mathrm{FOM}=|\mathrm{TCF}|~Q/10^6\sim207$~K$^{-1}$ (Table~S1, Supplementary Information, Section~2), one to two orders of magnitude higher than previously reported MEMS and lithium-niobate resonant thermal sensors. This shows that feedback can substantially reshape the temperature response beyond the intrinsic mechanical behavior.

This offers an alternative to conventional MEMS sensor design: rather than relying solely on material selection, structural geometry, or mechanical-mode engineering, temperature responsivity can instead be tuned post-fabrication through laser--cavity detuning and feedback-phase control. Combined with the spatially localized, multi-domain-coupled nature of the suspended resonator, this makes the approach attractive for integrated sensing, local temperature monitoring, and active stabilization of nearby photonic and MEMS components.

The same architecture also supports injection locking, trading the free-running state's environmental sensitivity for external-reference-governed stability, and could enable selective stabilization among competing regenerative modes in multimode systems. Taken together, these results establish the OEM feedback loop as an active, post-fabrication means of shaping both the temperature response and the frequency stability of mechanical oscillation, pointing toward reconfigurable sensing platforms and integrated optoelectromechanical frequency-control systems.

\section*{Acknowledgments}
This work was supported by the Defense Advanced Research Projects Agency (DARPA) under the OpTIm program (Grant No.\ HR00112320022); by the Department of Energy (DOE) under Award No.\ DE-AR0002067; and by DOE, Office of Science, Basic Energy Sciences (BES), under the Quantum Coherent Systems program (Contract No.\ DE-AC02-05CH11231). Device fabrication was performed at the Marvell Nanofabrication Laboratory at the University of California, Berkeley, and at the USC Core Center of Excellence in Nano Imaging. Work at the Molecular Foundry was supported by the Office of Science, Office of Basic Energy Sciences, of the U.S. Department of Energy under Contract No.\ DE-AC02-05CH11231.

\section*{Author contributions}
$\,\,\,$ M.Y. conceived the conceptual idea. R.Y. conducted the experiment, analyzed data, built the cavity delay model. Y.Y. fabricated the suspended chip. I.C. helped to conduct the open-loop phase measurement experiment. R.Y. drafted the manuscript, which was revised with significant input from M.Y., and contributions from all authors. M.Y. supervised the project.

\section*{Disclosures}
The authors declare no conflicts of interest.

\section*{Data Availability}
The data that support the findings of this study are available from the corresponding author upon reasonable request.

\section*{Supplemental document}
The Supplementary Information is included after the references.

\printbibliography
\end{refsection}

\onecolumn
\begin{refsection} 
\setcounter{section}{0}
\setcounter{figure}{0}
\setcounter{table}{0}
\setcounter{equation}{0}
\renewcommand{\thefigure}{S\arabic{figure}}
\renewcommand{\thetable}{S\arabic{table}}
\renewcommand{\theequation}{S\arabic{equation}}
\renewcommand{\theHsection}{SI.\arabic{section}}
\renewcommand{\theHfigure}{SI.\arabic{figure}}
\renewcommand{\theHtable}{SI.\arabic{table}}
\renewcommand{\theHequation}{SI.\arabic{equation}}
\captionsetup[table]{labelsep=period}
\begin{center}
{\LARGE\bfseries Supplementary Information}\\[0.6em]
{\large Giant thermal transduction via feedback-phase engineering in a thin-film lithium-niobate opto-electro-mechanical oscillator}
\end{center}
\vspace{0.5em}

\section{Comparison of the passive mechanical spectrum and open-loop gain}

\begin{figure}[htbp]
\centering
\includegraphics[width=0.48\linewidth]{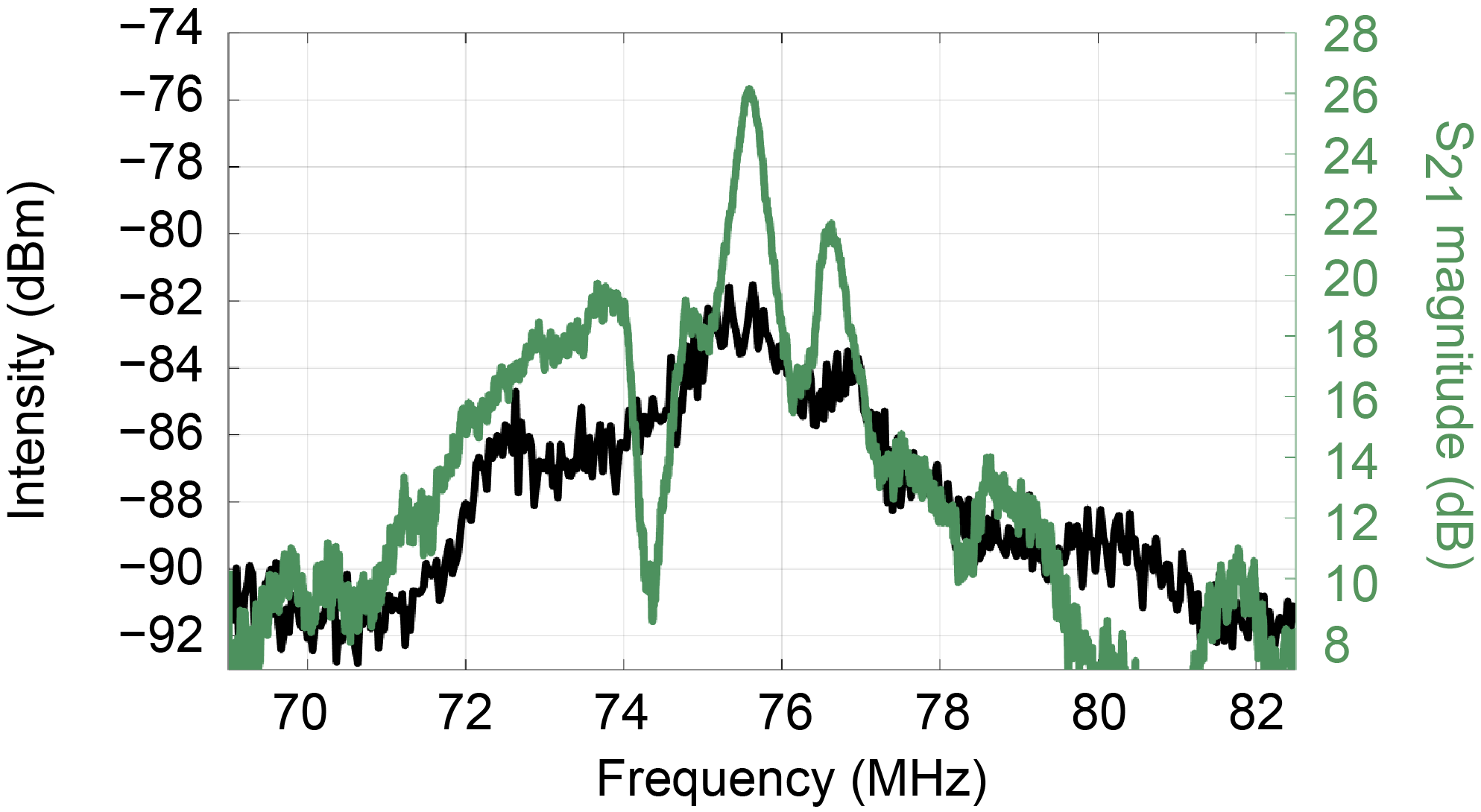}
\caption{\textbf{Comparison of the passive mechanical spectrum and open-loop gain.} Passive mechanical spectrum (black) and VNA-measured open-loop $S_{21}$ magnitude response (green). The peaks in the open-loop response occur near the passive mechanical resonances, indicating their contribution to the frequency-dependent loop gain. The two traces are plotted against separate vertical axes with different units and should not be compared in absolute amplitude.}
\label{fig:s1}
\end{figure}

To identify the mechanical contributions to the frequency-dependent loop gain, we compare the passive mechanical spectrum with the open-loop $S_{21}$ magnitude response measured using a vector network analyzer (VNA). As shown in Figure~\ref{fig:s1}, the prominent peaks in the open-loop response closely correspond to mechanical resonances observed in the passive spectrum. This agreement indicates that the mechanical resonances play a significant role in shaping the frequency-dependent magnitude response of the feedback loop.

\section{Evolution of the oscillator response beyond the enhanced-TCF regime}

\begin{figure}[htbp]
\centering
\includegraphics[width=0.98\linewidth]{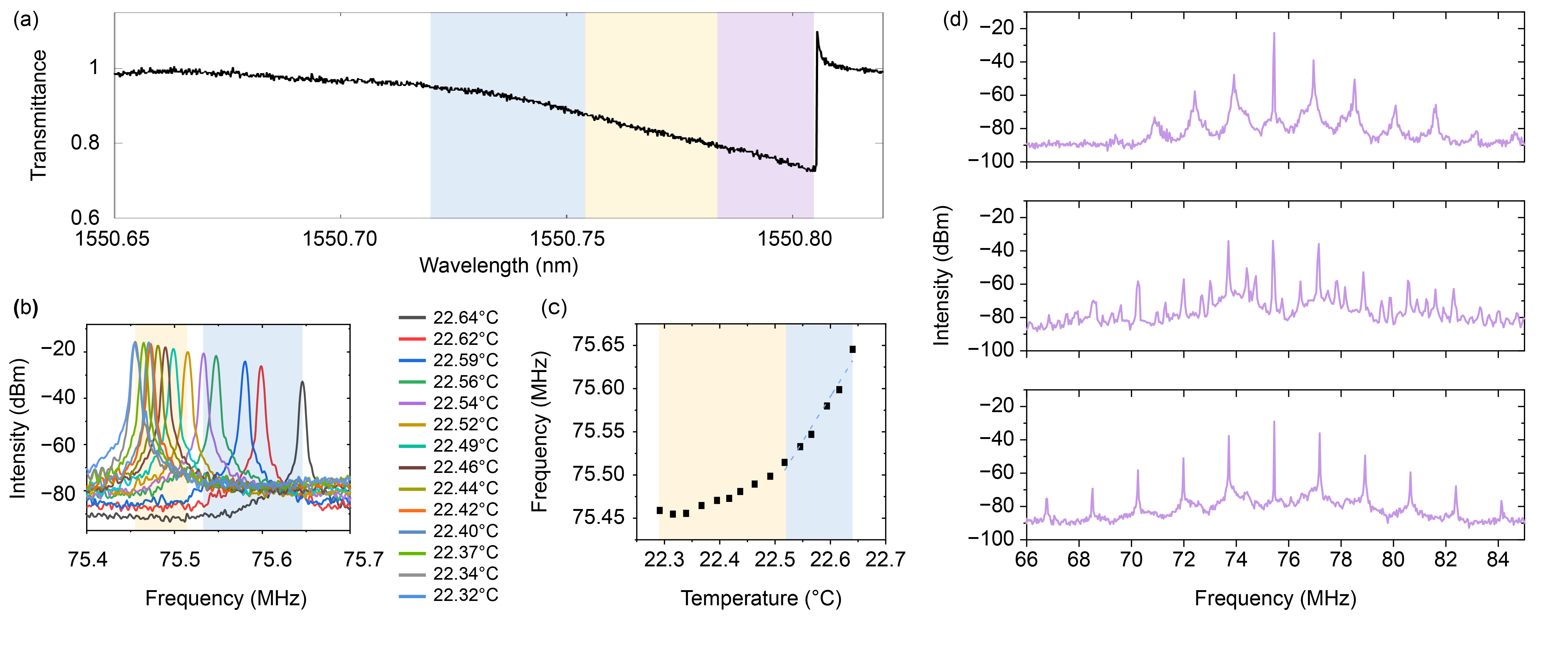}
\caption{\textbf{Oscillator behavior across the optical resonance.} (a) Optical transmission spectrum with three representative operating regimes highlighted: enhanced-TCF (blue), reduced-TCF single-frequency oscillation (yellow), and multifrequency oscillation (purple). (b) Closed-loop RF spectra measured at different temperatures in the enhanced-TCF regime. (c) Extracted oscillation frequencies as a function of temperature across the single-frequency operating regimes. (d) Representative multifrequency RF spectra in the third regime, where the spacing between adjacent peaks varies from approximately $0.8$ to $1.5$~MHz. The highlighted colors distinguish the experimentally observed operating regimes.}
\label{fig:s2}
\end{figure}

The enhanced temperature coefficient of frequency (TCF) for the electro-optomechanical (EOM) oscillator reported in the main text is observed over a limited laser--cavity detuning range. Figure~\ref{fig:s2}(a) highlights three distinct operating regimes on the measured optical transmission spectrum, each exhibiting different oscillation characteristics. At the on-chip optical power of approximately $3.5$~mW used in the experiment, the optical resonance exhibits a triangular lineshape due to thermal distortion. As the device temperature decreases at a fixed laser frequency, the cavity resonance blue shifts relative to the laser, driving the oscillator through three experimentally distinguishable regimes.

In the first regime, highlighted in blue, the feedback loop sustains a single dominant RF oscillation peak with the strongly enhanced temperature dependence discussed in the main text [Figure~\ref{fig:s2}(b)]. Further cooling brings the oscillator into a second regime, highlighted in yellow, where single-frequency oscillation is maintained but the frequency dependence on the externally controlled temperature becomes significantly weaker [Figure~\ref{fig:s2}(c)]. One possible explanation is that the increased intracavity optical power introduces stronger optomechanical backaction, which modifies the effective mechanical resonance frequency and may partially compensate for the temperature-induced frequency shift. Alternatively, enhanced intracavity optical heating may reduce the actual temperature variation of the suspended resonator relative to the temperature change indicated by the external controller. These possible mechanisms have not been independently quantified and warrant further investigation.

As the laser--cavity detuning decreases further, the oscillator enters a third regime, highlighted in purple, characterized by the emergence of multiple approximately equally spaced RF spectral peaks [Figure~\ref{fig:s2}(d)]. As the device temperature is further reduced and the optical resonance approaches the fixed laser frequency, the spacing between adjacent RF peaks varies from approximately $0.8$ to $1.5$~MHz. The emergence of this multifrequency response may be associated with nonlinear optomechanical interactions at higher intracavity optical power. However, the detailed mechanism underlying the RF comb formation remains to be investigated.

\clearpage

\subsection{Comparison with previously reported resonant temperature sensors}

Table~\ref{tab:si_tcf_q_review} places the measured TCF of the OEM feedback oscillator in the context of previously reported resonant MEMS and lithium-niobate (LN) temperature sensors, using the linewidth-normalized figure of merit $\mathrm{FOM}=10^{-6}|\mathrm{TCF}|Q$ introduced in the main text. The OEM feedback oscillator reaches an FOM of approximately $207$~K$^{-1}$, one to two orders of magnitude higher than the other entries in the table.

Among the LN-based entries, a related report of a comparable TCF ($\sim2\times10^4$~ppm/K) in a different, suspended optomechanical LN platform \cite{High_harmonic_optomechanical_oscillations_LN} briefly attributes this enhancement to oxide-layer properties despite the device being suspended -- a mechanism not expected to dominate thermal transduction in a suspended geometry -- without presenting experimental evidence (e.g., independent temperature-dependent or phase measurements) identifying the physical origin of the effect; this value is marked with the superscript $^{\mathrm{U}}$ in Table~\ref{tab:si_tcf_q_review}. By contrast, the giant TCF reported in the present work is corroborated by independent temperature-dependent and open-loop phase measurements and is quantitatively reproduced by the detuning-dependent delay model of Section~\ref{sec:si_phase_model}, directly tying the observed enhancement to feedback-phase selection via the laser--cavity detuning.

\begin{table}[htbp]
\centering
\caption{\textbf{Temperature coefficients and quality factors of representative MEMS and lithium-niobate devices.}}
\label{tab:si_tcf_q_review}
\renewcommand{\arraystretch}{1.10}
\resizebox{\textwidth}{!}{%
\begin{tabular}{@{}llcccc@{}}
\toprule
\textbf{Device and reference} & \textbf{Sensing mode} & \textbf{TCF (ppm/K)} & $\boldsymbol{Q_{\mathrm{res}}}$ & $\boldsymbol{Q_{\mathrm{osc}}}$ & \textbf{FOM (K$^{-1}$)} \\
\midrule
\multicolumn{6}{l}{\textit{Other MEMS resonant thermometers}} \\
\addlinespace[2pt]
Bare SMP near $T_g$ ($45~^\circ$C) \cite{polymer_8e4ppmK} & P & $\sim-8\times10^4$ & $\sim2$ & NA & $\sim0.160^{\mathrm{P}}$ \\
Bare SMP at $30~^\circ$C \cite{polymer_8e4ppmK} & P & $\sim-1.2\times10^4$ & $34$ & NA & $\sim0.408^{\mathrm{P}}$ \\
SiN$_x$/SMP at $30~^\circ$C \cite{polymer_8e4ppmK} & P & $\sim-4\times10^3$ & $1.0\times10^3$ & NA & $\sim4.20^{\mathrm{P}}$ \\
Stressed CMOS--MEMS beam \cite{clamp_4537ppmK} & NA & $-4.5\times10^3$ & NA & NA & NA \\
CMOS--MEMS feedback sensor \cite{clamp_-3550ppmK} & O & $-3.6\times10^3$ & NA & NA & NA \\
Graphene--Au resonator \cite{graphene_-3180ppmK} & P & $\sim-3.2\times10^3$ & NA & NA & NA \\
AlN-on-Si TETF oscillator \cite{tu2022coupled_out_of_plane_vib_1000ppmK} & O & $-1.1\times10^3$ & $670$ & NA & NA \\
High-resolution MEMS thermometer \cite{high_res_MEMs} & P/O & $-866.84^{\mathrm{P}}$ & $4.8\times10^4$ & NA & $41.8^{\mathrm{P}}$ \\
Dual-mode differential thermometer \cite{dual_mode_1480ppmK} & NA (beat) & $+1.5\times10^3$ & NA & NA & NA \\
Internal-resonance comb thermometer \cite{comb_space_-11481ppmK} & P (comb) & $-1.2\times10^4$ & NA & NA & NA \\
TETF-pair beat thermometer \cite{BEATING_SIGNAL_9929ppmK} & NA (beat) & $|\mathrm{TCF}|\sim10^4$ & NA & NA & NA \\
\midrule
\multicolumn{6}{l}{\textit{Lithium-niobate acoustic and optomechanical devices}} \\
\addlinespace[2pt]
LN/SiO$_2$ S$_0$ resonator ($-50^\circ$) \cite{TFLN_on_SiO_28ppmK} & P & $+27.7$ & $132$ & NA & $0.00366^{\mathrm{P}}$ \\
LN SAW thermometer \cite{LN_SAW_thermometer} & P & $-81.1$ & $517$ & NA & $0.0419^{\mathrm{P}}$ \\
LN microdisk oscillator \cite{LN_optical_lasing_qianglin} & O & NA & $5.4\times10^3$ & $2.16\times10^6$ & NA \\
LN thickness mode, 4 K \cite{thickness_mode_optomechanical_sensor_LN} & P & NA & $1.2\times10^4$ & NA & NA \\
LN SH$_0$ LVR, mode A \cite{Eisner2021_LN_TCF} & P & $-95.27$ & $415$ & NA & $0.0395^{\mathrm{P}}$ \\
LN SH$_0$ LVR, mode B \cite{Eisner2021_LN_TCF} & P & $-95.43$ & $508$ & NA & $0.0485^{\mathrm{P}}$ \\
LN SAW near-IR detector, bare control \cite{SAW_IR_detector} & P & $\sim-165$ & $400$ & NA & $\sim0.0661^{\mathrm{P}}$ \\
TFLN Lamb-wave IR sensor \cite{lamb_wave_IR_sensor} & P & NA & NA & NA & NA \\
LN phononic-crystal MIR oscillator \cite{Photonic_crystal_LN_MIR_sensor} & O & NA & $1.0\times10^3$ (loaded) & NA & NA \\
LN-film SAW, Rayleigh \cite{LN_SAW_-66} & P & $-66$ & $526$ & NA & $0.0347^{\mathrm{P}}$ \\
LN nanocavity \cite{High_harmonic_optomechanical_oscillations_LN} & P & $\sim+2\times10^4$ $^{\mathrm{U}}$ & $65.38$ & NA & $\sim1.31^{\mathrm{P}}$ \\
TFLN optomechanical sensor \cite{uncooledlownoisethinfilmoptomechanical} & P & $-124^{\mathrm{P}}$ & $1.1\times10^3$ & $5.1\times10^4$ & $0.139^{\mathrm{P}}$ \\
\midrule
\textbf{This work: OEM loop} & \textbf{O} & $\boldsymbol{+1.24\times10^4}$ & $\boldsymbol{210}$ & $\boldsymbol{1.67\times10^4}$ & $\boldsymbol{207}^{\mathrm{O}}$ \\
\bottomrule
\end{tabular}%
}
\vspace{3pt}
\caption*{The linewidth-normalized temperature-responsivity figure of merit is $\mathrm{FOM}=10^{-6}|\mathrm{TCF}|Q$ (K$^{-1}$), with TCF expressed in ppm/K. $Q_{\mathrm{res}}$ is the reported mechanical resonator quality factor, including loaded or electrical-resonance $Q$ values; $Q_{\mathrm{osc}}$ is the free-running spectral oscillation effective $Q$ values. The relevant sensing regime determines which $Q$ is used. P: passive or externally driven sensing; O: self-sustained oscillation. P/O denotes both; $^{\mathrm{U}}$ TCF value as claimed by the authors, with no experimental data shown to explain the cause. Beat/comb sensing requires the linewidth of that readout, rather than a constituent mode. NA: unavailable, unverified, or non-comparable.}
\end{table}

\clearpage

\section{Transfer function from mechanically induced detuning modulation to intracavity optical energy modulation}
\label{sec:si_phase_model}

Mechanical displacement shifts the optical resonance frequency of the suspended resonator, producing a time-dependent laser--cavity detuning at the mechanical oscillation frequency $f_{\rm osc}$. This detuning modulation drives an intracavity optical energy modulation whose amplitude and phase depend on the operating detuning and modulation frequency. Here, we derive the corresponding small-signal transfer function from the temporal coupled-mode equation, relate the intracavity response to the transmitted optical signal, and use its relative phase to obtain the cavity contribution to the feedback frequency-selection condition.

For a single optical cavity mode, the intracavity field amplitude $a(t)$ obeys the temporal coupled-mode equation \cite{optomechanic_equation}
\begin{equation}
\frac{da}{dt}=\left[i\Delta(t)-\frac{\kappa}{2}\right]a
+\sqrt{\kappa_{\rm ex}}\,s_{\rm in},
\label{eq:si_cmt}
\end{equation}
where $\Delta(t)=\omega_L-\omega_c(t)$ is the instantaneous laser--cavity detuning, $\kappa=\kappa_0+\kappa_{\rm ex}$ is the total optical energy-decay rate, and $\kappa_0$ and $\kappa_{\rm ex}$ are the intrinsic and external coupling rates, respectively. The intracavity field is normalized such that $U(t)=|a(t)|^2$ is the intracavity optical energy and $|s_{\rm in}|^2$ is the input optical power. We write $\kappa=2\pi\Gamma$, where $\Gamma$ is the optical linewidth (FWHM) in hertz; detuning and optical decay rates in Eq.~\eqref{eq:si_cmt} are in angular-frequency units.

For a constant operating detuning $\Delta$, the steady-state solution is
\begin{equation}
\begin{aligned}
a_0(\Delta)&=\frac{\sqrt{\kappa_{\rm ex}}\,s_{\rm in}}{\kappa/2-i\Delta},\\
U_0(\Delta)&=|a_0(\Delta)|^2
=\frac{\kappa_{\rm ex}|s_{\rm in}|^2}{(\kappa/2)^2+\Delta^2}.
\end{aligned}
\label{eq:si_steady_state}
\end{equation}

A small mechanical displacement produces a detuning perturbation $\delta\Delta(t)$ around this operating point, so that $\Delta(t)=\Delta+\delta\Delta(t)$ and $a(t)=a_0+ \delta a(t)$. Substitution into Eq.~\eqref{eq:si_cmt}, followed by subtraction of the steady-state equation, gives
\begin{equation}
\frac{d\delta a}{dt}
=\left(i\Delta-\frac{\kappa}{2}\right)\delta a
+i a_0 \delta\Delta(t)+i \delta\Delta(t)\delta a(t).
\label{eq:si_perturbed_field_exact}
\end{equation}
Since $\delta a$ is itself first order in $\delta\Delta(t)$, the product $\delta\Delta(t)\delta a(t)$ is second order in the small perturbation. Neglecting this term yields the linear small-signal equation
\begin{equation}
\frac{d\delta a}{dt}
=\left(i\Delta-\frac{\kappa}{2}\right)\delta a+i a_0 \delta\Delta(t).
\label{eq:si_linearized_field}
\end{equation}

Using the Fourier convention $\delta\Delta(t) = \delta\Delta_{f_{\rm osc}} e^{-i2\pi f_{\rm osc}t}+\delta\Delta_{f_{\rm osc}} e^{i2\pi f_{\rm osc}t}$ and $\delta a(t)=\delta a_{f_{\rm osc}}e^{-i2\pi f_{\rm osc}t}+\delta a^*_{f_{\rm osc}}e^{i2\pi f_{\rm osc}t}$, the Fourier components of the field perturbation and its complex conjugate satisfy
\begin{equation}
\begin{aligned}
\delta a_{f_{\rm osc}}
&=\frac{i a_0\,\delta\Delta_{f_{\rm osc}}}{\kappa/2-i\Delta-i2\pi f_{\rm osc}},\\
(\delta a^*)_{f_{\rm osc}}
&=\frac{-i a_0^*\,\delta\Delta_{f_{\rm osc}}}{\kappa/2+i\Delta-i2\pi f_{\rm osc}}.
\end{aligned}
\label{eq:si_sidebands}
\end{equation}
The intracavity energy, expanded to first order in $\delta a$, is $U(t)=U_0+a_0^*\delta a(t)+a_0\delta a^*(t)$, so its modulation component is
\begin{equation}
\delta U_{f_{\rm osc}}
=a_0^*\delta a_{f_{\rm osc}}+a_0(\delta a^*)_{f_{\rm osc}}.
\label{eq:si_energy_mod}
\end{equation}
Inserting Eq.~\eqref{eq:si_sidebands} into Eq.~\eqref{eq:si_energy_mod} and collecting terms yields
\begin{equation}
\begin{aligned}
\delta U_{f_{\rm osc}}
&=\frac{-2\Delta\,U_0(\Delta)\,\delta\Delta_{f_{\rm osc}}}
{(\kappa/2-i2\pi f_{\rm osc})^2+\Delta^2}.
\end{aligned}
\label{eq:si_energy_response}
\end{equation}
The complex transfer function from mechanically induced detuning modulation to intracavity optical energy modulation is therefore shown in Eq.~\eqref{eq:si_GU}. At finite $f_{\rm osc}$ and $\Delta$, the complex denominator produces a phase shift that varies with both detuning and modulation frequency.

\begin{equation}
\begin{aligned}
H_U(\Delta,f_{\rm osc})
&\equiv\frac{\delta U_{f_{\rm osc}}}{\delta\Delta_{f_{\rm osc}}} =\frac{-2\Delta U_0(\Delta)}
{(\kappa/2-i2\pi f_{\rm osc})^2+\Delta^2}.
\end{aligned}
\label{eq:si_GU}
\end{equation}

\section{Estimation of the feedback-loop phase-cycle index from the open-loop phase spectrum}
\label{sec:si_loop_delay}

\begin{figure}[h]
\centering
\includegraphics[width=0.48\linewidth]{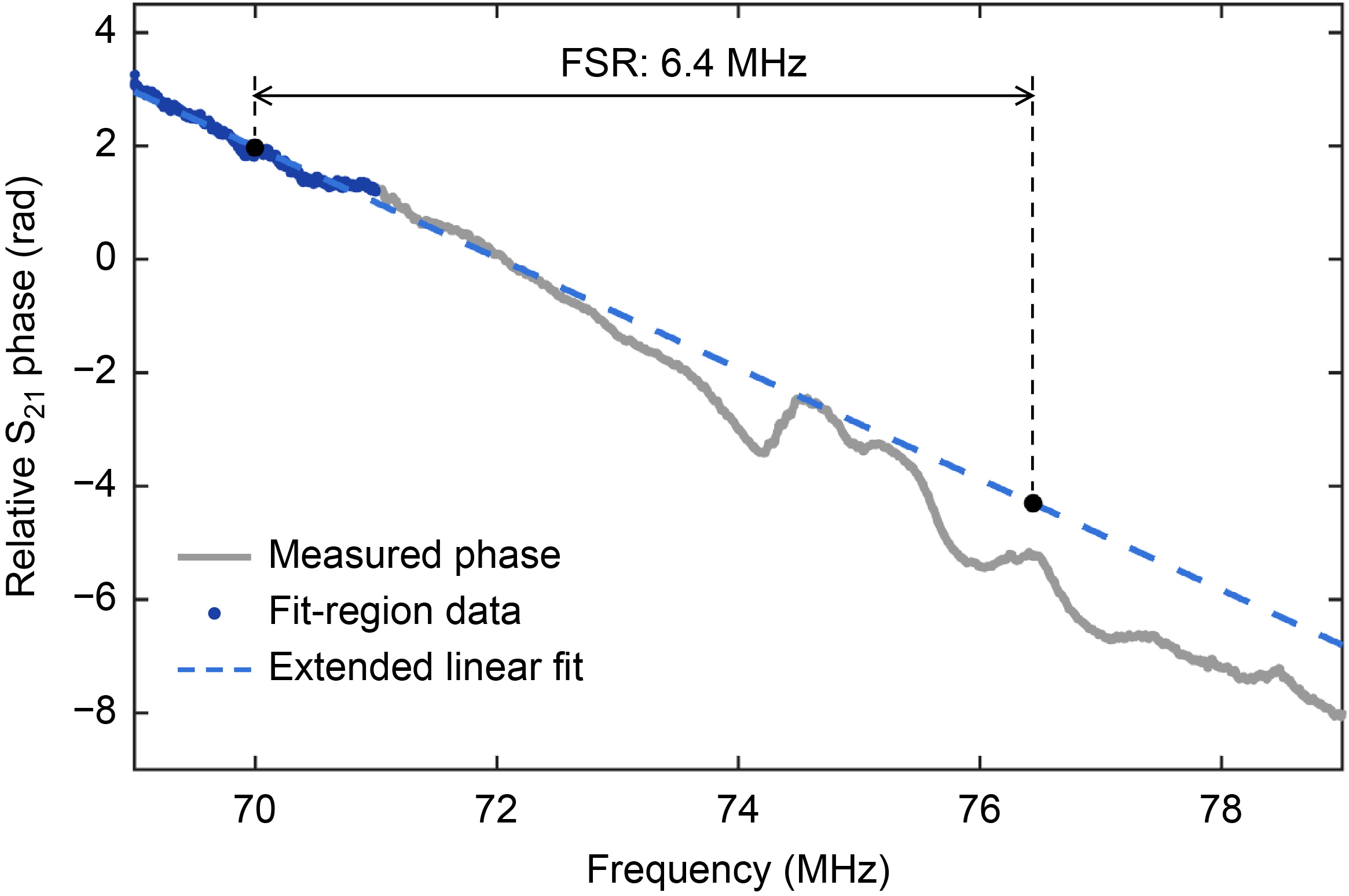}
\caption{Extraction of the loop FSR from the open-loop $S_{21}$ phase spectrum. The gray curve represents the measured phase response, while the blue dashed line shows the linear fit to the off-resonance region (69--71~MHz) extrapolated over the measured frequency range. The frequency separation corresponding to a $2\pi$ phase change yields a loop FSR of approximately $6.4$~MHz.}
\label{fig:si_fsr}
\end{figure}

To estimate the feedback-loop phase-cycle index $m$ used in the main-text model, we analyze the measured open-loop $S_{21}$ phase spectrum shown in Figure~\ref{fig:si_fsr}. To avoid localized phase variations associated with mechanical resonances, we perform a linear fit to the phase response between 69 and 71~MHz and extrapolate the fitted line over the measured frequency range. The frequency interval corresponding to a $2\pi$ phase change yields a loop free spectral range of approximately $\mathrm{FSR}_{\rm loop}=6.4$~MHz.

The oscillation frequency can be expressed in terms of the loop FSR as
\begin{equation}
f_{\rm osc}=m\,\mathrm{FSR}_{\rm loop}+f_{\rm off},
\label{eq:si_mode_index}
\end{equation}
where $m$ is the integer feedback-loop phase-cycle index and $f_{\rm off}$ represents the frequency offset associated with the residual loop phase. Adopting the convention $0\leq f_{\rm off}<\mathrm{FSR}_{\rm loop}$, the measured oscillation frequency at the calibration point for the model of $73.2$~MHz gives
\begin{equation}
m=\left\lfloor\frac{f_{\rm osc}}{\mathrm{FSR}_{\rm loop}}\right\rfloor=\left\lfloor\frac{73.2}{6.4}\right\rfloor=11,
\label{eq:si_mode_selection}
\end{equation}
with $f_{\rm off}=73.2-11\times6.4=2.8$~MHz. We therefore adopt $m=11$ as the effective phase-cycle index in the cavity effective delay model in the main text. This assignment is consistent with the measured S$_{21}$ loop FSR and the chosen frequency-offset convention.

\begin{figure}[h]
\centering
\includegraphics[width=0.48\linewidth]{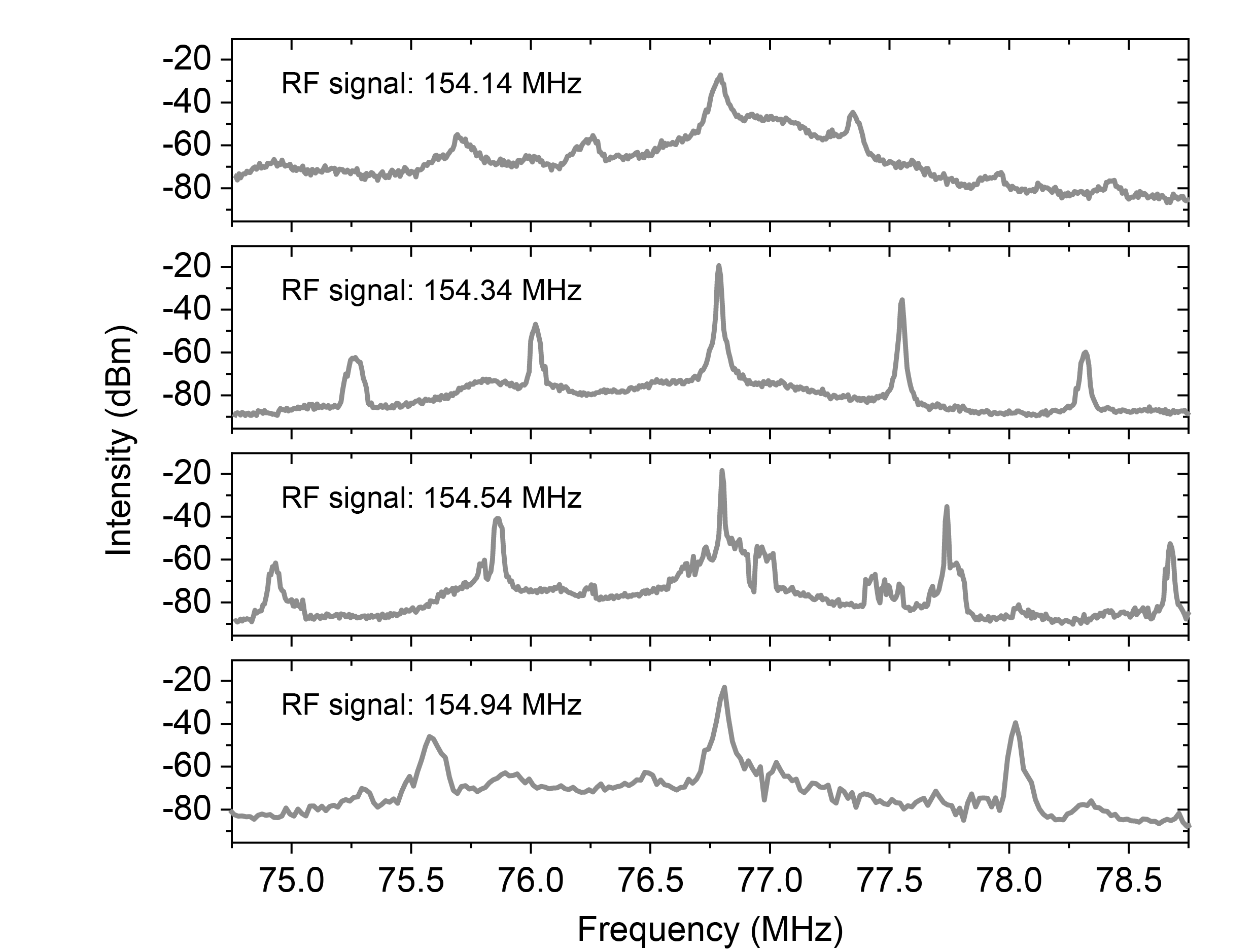}
\caption{\textbf{Mechanical frequency-comb generation in the unlocked injection regime.} Measured RF spectra of the opto-electro-mechanical oscillator under external injection at driving frequencies of 154.14, 154.34, 154.54, and 154.94 MHz (from top to bottom). Outside the injection-locking range, the oscillator exhibits multiple spectral peaks characteristic of mechanical frequency-comb generation. The comb structure and spacing vary with the external driving frequency, demonstrating that the unlocked oscillation dynamics can be controlled through the external RF excitation.}
\label{fig:s4}
\end{figure}

\section{Multifrequency oscillation outside the external injection-locking range}
The main text establishes the $2{:}1$ subharmonic injection-locking regime, in which the mechanical oscillation frequency follows half the externally applied RF modulation frequency. Here, we consider the spectral response outside the range over which stable single-frequency locking is observed. Unlike the temperature-dependent multifrequency regime described in Section~2, these measurements are obtained by changing the external modulation frequency in the injection-locking configuration.

When the modulation frequency moves away from the locking region, the RF response develops multiple spectral peaks rather than remaining concentrated at a single entrained oscillation frequency. Representative measurements at external drive frequencies of $154.14$, $154.34$, $154.54$, and $154.94$~MHz exhibit multifrequency spectra whose peak spacing varies with the applied drive frequency (Figure~\ref{fig:s4}). These observations show that the externally driven response can remain spectrally structured even when the oscillator is no longer locked to the reference.

A possible interpretation is that the external modulation perturbs the regenerative oscillator while its preferred free-running frequency remains determined by the loop response. However, the present measurements do not independently determine the nonlinear coupling mechanism, establish mutual phase coherence among the observed components, or distinguish all possible origins of the spectral splitting. We therefore describe these results as externally driven multifrequency or comb-like spectra, without asserting a specific mechanical-frequency-comb generation mechanism.

\printbibliography[title={Supplementary References}]
\end{refsection}
\end{document}